\documentclass[preprint,prd,amsmath,amssymb,nobibnotes,nofootinbib]{revtex4}
\usepackage{amsfonts}
\usepackage{revsymb}
\usepackage{graphicx,epsfig}
\usepackage{placeins}

\begin{document}
\title {Massive-field approach to the scalar self force in curved spacetime}
\author{Eran Rosenthal}
\affiliation{Department of Physics, University of Guelph, Guelph, Ontario,
Canada N1G 2W1}
\date{\today}

\begin{abstract}
We derive a new regularization method for the calculation of the
(massless) scalar self-force in curved spacetime. In this method,
the scalar self-force is expressed in terms of the difference
between two retarded scalar fields: the massless scalar field, and
an auxiliary massive scalar field. This field difference combined
with a certain limiting process gives the expression for the
scalar self-force. This expression provides a new self-force
calculation method.
\end{abstract}

\maketitle

\section{Introduction and summary}

The motion of an electric charged point-like object in a fixed
background spacetime, is affected by the coupling between the
object's own charge, and the field that this charge induces. This
coupling results in a {\em self force} (also known as the "radiation
reaction force") acting on the object.  At leading order, the
object's acceleration due to this self force  (in the absence of
non-gravitational external interactions) is proportional to
$q^2/\mu$, where $q$ and $\mu$ denotes the object's charge and
mass, respectively. This leading order is obtained by treating the
particle's field as a linear perturbation over a fixed curved
background spacetime. Analogous to the electromagnetic self-force
there are other types of self forces: a scalar self-force is
induced by a scalar charge, and a gravitational self-force is
induced by the object's mass (in this case the object's
acceleration due to the gravitational self-force is proportional
to $\mu$).

In recent years, there has been growing interest in the
calculation of the self force in curved spacetime \cite{MST}-
\cite{DW}. Formal expressions of the self force have been derived:
Mino, Sasaki and Tanaka \cite {MST}, and independently Quinn and
Wald \cite{QW}, have recently obtained an expression for the
gravitational self force; previously, DeWitt and Brehme \cite{DB}
 obtained an expression for the electromagnetic self force;
and Quinn \cite{Quinn} recently obtained an expression for the
scalar self force. In the case of a weak gravitational field,
explicit self-force expressions were found by DeWitt and DeWitt
\cite{2De}, and by Pfenning an Poisson \cite{PP}. Analysis of the
self force in curved spacetime also has a practical motivation:
one possible source for LISA -- the planned space-based
gravitational wave detector \cite {LISA}, is a binary system with
an extreme mass ratio, which inspirals toward coalescence. Here,
the self force is required for the calculation of the accurate
orbital evolution of such systems. These orbits are needed in
order to design templates for the gravitational waveforms of the
emitted gravitational radiation. A calculation method for the self
force in such binary systems, was devised by Barack and Ori
\cite{BO}, this method was recently improved
\cite{BOSCALAR,BMNOS}, and also implemented numerically in certain
cases \cite{Burkoim1,Burkoim2}. For other approaches to the self
force problem see \cite{Lousto,Detweiler,NS}.

Previous analyses have provided several expressions for the scalar
self force, which are equivalent to each other. In Quinn's
derivation \cite{Quinn}, the expression for the scalar self force
is composed of two different types of terms: explicit local terms;
and a non-local term, expressed as a certain integral over (the
partial derivatives of) the retarded scalar Green's function. We
shall refer to these
terms as the local terms, and the non-local term, respectively.
Very recently, Detweiler and Whiting \cite{DW}, developed a different
method to express the self force in curved spacetime. In the
scalar-field variant of their analysis, they showed that the scalar self
force can be obtained from $\psi^R$, which is a certain
non-retarded solution of the homogeneous scalar-field equation.

In calculating the self force acting on a point like object, one
typically encounters a divergent expression, from which the
 finite (and correct) expression for the self force should be
 obtained using a certain {\em regularization} method.
 In this manuscript we present
a new self-force regularization method. We consider a point-like
scalar charge in curved spacetime, which induces a scalar field
$\phi$. We show (under certain assumptions) that the scalar self
force acting on this scalar charge can be expressed in terms of
two retarded scalar fields, which satisfy prescribed differential
equations. These fields are: $\phi$ -- which satisfies the
inhomogenous massless scalar field equation, with a charge density
$\rho$; and $\phi_m$ -- which satisfies the inhomogenous {\em
massive} scalar field equation, with exactly the same charge
density $\rho$ (here $m$ denotes the mass of the massive scalar
field). More specifically, we show that the scalar self force is
given by
\begin{equation}\label{sf}
f_\mu^{self}(z_0)=q\lim_{m\rightarrow\infty}\left\{
\lim_{\delta\rightarrow 0} \Delta\phi_{,\mu}(x)+
\frac{1}{2}q\biglb[m^2n_\mu(z_0)+ma_\mu(z_0)\bigrb]\right\}\,,
\end{equation}
where,
\begin{equation}\label{deltaphi}
 \Delta\phi(x)\equiv\phi(x)-\phi_m(x)\,\,.
\end{equation}
Here, $z_0$ is the self-force evaluation point on the object's
world line, and $x$ is a point near the word line, defined as
follows. At $z_0$ we construct a unit spatial vector $n^\mu$,
which is perpendicular to the object's world line but is otherwise
arbitrary (i.e. at $z_0$ we have $n^\mu n_\mu=1$ , $n^\mu
u_\mu=0$). In the direction of this vector we construct a
geodesic, which extends out an invariant length $\delta$ to the
point $x(z_0,n^\mu,\delta)$; throughout this manuscript $u^\mu$
and $a^\mu$ denote the object's four-velocity and four-acceleration, at $z_0$, respectively.

We show below that the value of $f_\mu^{self}$ given by Eq.
(\ref{sf}) is identical to the scalar self force obtained by Quinn
\cite{Quinn}. Therefore, the problem of calculating the scalar
self force in curved spacetime is equivalent to the problem of
solving the two scalar partial differential equations for the
fields $\phi$ and $\phi_m$, and then carrying out the prescribed
set of operations summarized by Eq. (\ref{sf}). Note that for
the massive field $\phi_m$, only the asymptotic behavior as
$m\rightarrow\infty$ is required.

We mention here a similar method developed by Coleman
\cite{Coleman} for the derivation of the electromagnetic self
force in flat spacetime (the Abraham-Lorentz-Dirac term
\cite{Abraham,Lorentz,Dirac}). In this method, the self-force
regularization is obtained by replacing the Green's function of the
electromagnetic four-potential with a different (more regular)
function, which depends on a certain parameter. This new function
behaves like the original electromagnetic Green's function at the
limit where the parameter approaches infinity. Coleman obtained
this new function from the Green's function of a fictitious massive
electromagnetic field. This regularization method is similar to
Pauli-Villars regularization in quantum field theory \cite{PV}.

The derivation of Eq. (\ref{sf}) is based on several properties of
the fields $\phi$ and $\phi_m$, which we now briefly summarize.
Previous methods \cite{Quinn,BMNOS,BOSCALAR,MNS} used the
following decomposition of the scalar field near the object's
world line: $\phi(x)=\phi_{dir}(x)+\phi_{tail}(x)$.
 In this decomposition the
direct field $\phi_{dir}(x)$ is a field that propagates along
the future light cone emanating from the object. Near the object's
world line this direct field can be determined using a local
analysis \cite{MNS}, and it diverges as $\delta\rightarrow0$. The
second field in this decomposition, the tail field
$\phi_{tail}(x)$, is of nonlocal nature, i.e., its value
generically depends on the entire past history of the object's
world line. Moreover, the tail field is finite at the limit
$\delta\rightarrow0$, and it vanishes in flat spacetime (in 1+3
dimensions). This tail field was found to be useful in the
calculation of the self force in curved spacetime \cite{BOSCALAR}
-- the above mentioned nonlocal term of the self force, can be
directly obtained from $\phi_{tail,\mu}$ near the object's world
line.

In a similar manner, near the object's world line the massive
field $\phi_m(x)$ can be decomposed according to
$\phi_m(x)=\phi_{m(dir)}(x)+\phi_{m(tail)}(x)$. Here again, the
massive direct field $\phi_{m(dir)}$ is a field that is
propagated along the future light cone emanating from the object,
and it diverges as $\delta\rightarrow0$. The massive tail field
$\phi_{m(tail)}$, is a field of a non-local nature, and is
non-singular on the world line (for a finite value of $m$).
However, unlike the massless tail field, this massive tail field
does not vanish in flat spacetime.

The formal expressions for the two direct fields $\phi_{m(dir)}$
and $\phi_{dir}$ (see below) show that they are identical. This
property is an important element in our construction, since it
implies that by subtracting $\phi_{m}$ from $\phi$ [see Eq.
(\ref{deltaphi})] we obtain
\begin{equation}\label{del2}
\Delta\phi(x)=\phi_{tail}(x)-\phi_{m(tail)}(x)\,,
\end{equation}
which is finite as $\delta\rightarrow0$, while the individual
fields $\phi$ and $\phi_m$ diverge at this limit. Similarly, the
partial derivatives $\Delta\phi_{,\mu}$ are also finite at this
limit. However, their values at this limit depend on the direction
along which the limit $\delta\rightarrow0$ is taken (i.e., they
depend on $n_\mu$). This directional dependency is completely
annihilated by the term $(qm^2/2)n_\mu$ in Eq. (\ref{sf}).
Therefore, the expression obtained by taking the limit
$\delta\rightarrow0$ in this equation does not depend on $n_\mu$.
To illustrate some of these properties consider the following
simplest example. Consider a static unit point charge situated at
$\vec{z}_0$, in a flat spacetime. In this case, the fields $\phi$
and $\phi_m$, which satisfy equations (\ref{scalar}) and
(\ref{scalarm}), respectively, are given by
\[
\phi=\phi_{dir}=\frac{1}{r}\,\,,\,\,\phi_m=\frac{1}{r}+\frac{1}{r}
(-1+e^{-mr})\,.
\]
Here $r\equiv|\vec{x}-\vec{z}_0|$. We therefore find that
$\Delta\phi=-\phi_{m(tail)}=\frac{1}{r}(1-e^{-mr})$, and the
gradient of this expression in the vicinity of the particle is
$\nabla(\Delta\phi)=-\frac{m^2}{2}{\hat{r}}+O(r)$, where $\hat{r}$
is a radial unit vector.
 Hence, both
$\Delta\phi$ and its gradient remain finite at the limit
$\delta\rightarrow 0$. Moreover, in this case the self force
obviously vanishes -- a result that is easily derived from Eq.
(\ref{sf}).

The calculation of the scalar self force, naturally involves
derivatives of the field -- ${\Delta\phi}_{,\mu}$ in our
construction. By virtue of (\ref{del2}), these derivatives will
include both $\phi_{tail,\mu}$ and $\phi_{m(tail),\mu}$. As we
mentioned above, the non-local part of the self force can be
obtained from $\phi_{tail,\mu}$. Therefore in our method
$\phi_{m(tail),\mu}$, should give rise to the (above mentioned)
local terms of the self force\footnote{We comment that there is
also a
 certain local contribution to the self-force
 coming from $\phi_{tail,\mu}$; this local contribution is
 taken into account in the calculations below.}. This might look strange at first
sight, because
 the derivatives $\phi_{m(tail),\mu}$ depend on the entire
past history of the object world line. Note, however that in our
method the self force is obtained by taking the limit
$m\rightarrow\infty$ in Eq. (\ref{sf}). At this limit the
contribution to Eq. (\ref{sf}) from the term $\phi_{m(tail),\mu}$
 is a local one (see calculations below).
 This behavior may be traced to the properties
 of the massive scalar Green's function at the limit $m\rightarrow\infty$.

 In Sec. \ref{massapp} we present the massive-field approach, and
show that $f_\mu^{self}$ which is given by Eq. (\ref{sf}) is
equivalent to the standard expression for the scalar self force,
given in \cite{Quinn}.

%%%%%%%%%%%%%%%%%%%%%%%%The massive field approach%%%%%%%%%%%%%%%%%%%%%%%%
\section{The massive field approach}\label{massapp}

 We consider a scalar field $\phi(x)$, which satisfies the scalar
field equation
\begin{equation}\label{scalar}
\Box\phi=-4\pi\rho\, .
\end{equation}
 Here $\Box\phi\equiv{\phi_{;\mu}}^\mu$, and $\rho(x)$
is the charge density of the scalar object. We use the signature
(-+++), and natural units where $c=1$ throughout. Our purpose is
to derive an expression for the self force on a point-like object.
We therefore consider a charge density of a point particle. The
particle's world line is denoted by $z(\tau)$, where $\tau$ is the
particle's proper time (we allow an arbitrary acceleration --
presumably caused by an arbitrary external force acting on the
particle), the particle's scalar charge is denoted by $q$ and the
particle's charge density is therefore
\begin{equation}\label{density}
\rho(x)=q\int_{-\infty}^{\infty}\frac{1}{\sqrt{-g}}
\delta^4[x-z(\tau)]d\tau \,.
\end{equation}
Here $g(x)$ is the determinant of the background metric. Next, we
define the field of the scalar force to be \footnote{This
definition conforms with definitions given in \cite{Quinn,BO}; a
different definition is also possible \cite{MNS}.}
\begin{equation}\label{sforce}
F_{\mu}(x)\equiv q\phi_{,\mu} \, .
\end{equation}
This field diverges as $x$ approaches the particle's world line.
In order to extract the self force from this singular field, we
need to regularaize this expression. For this purpose, we
introduce an auxiliary field $\phi_m(x)$ which satisfies the
massive field (Klein-Gordon) equation\footnote{Note that here the
field's mass $m$ has the dimensions of $(length)^{-1}$.}
\begin{equation}\label{scalarm}
(\Box-m^2)\phi_m=-4\pi\rho\, .
\end{equation}
Here the charge density $\rho$ is the same charge density introduced
above [i.e., it is given by Eq. (\ref{density})]. First we
consider the field's mass $m$ to be some fixed (large) quantity.
Similar to Eq. (\ref{sforce}) we define the field of the massive
  scalar force to be
\begin{equation}\label{smforce}
F_{(m)\mu}(x)\equiv q{\phi_{m}}_{,\mu}.
\end{equation}
Like the massless field $F_\mu$, the massive field $F_{(m)\mu}$
diverges as $x$ approaches the particle's world line. However, as
we show below, their difference remains finite at this limit.
 We therefore construct the difference field $\Delta\phi$ given by Eq. (\ref{deltaphi}),
and introduce the difference between the massless and the massive
scalar forces
\begin{equation}\label{deltafmu}
{\Delta F}_{\mu}(x)\equiv
F_{\mu}(x)-F_{(m)\mu}(x)=q\Delta\phi_{,\mu}\,.
\end{equation}
This field is an essential element in our construction, since in
our method the scalar self force is calculated from ${\Delta
F}_{\mu}$ -- see Eq. (\ref{sf}). To analyze the properties of this
field, we derive a formal expression for its value near the
particle's world line. This expression will later be used to
obtain Eq. (\ref{sf}).

First, we derive a formal expression for the field
${\Delta\phi}$ using the corresponding  retarded Green
functions of the massive, and massless fields. The retarded
solutions of equations (\ref{scalar},\ref{scalarm}) for the charge
density given by Eq. (\ref{density}), are formally given by
\begin{eqnarray}\label{phix}
\phi(x)&=&q\int_{-\infty}^{\infty}G(x|z(\tau))d\tau,\\\label{phixm}
\phi_m(x)&=&q\int_{-\infty}^{\infty}G_m(x|z(\tau))d\tau\,.
\end{eqnarray}
Here, $G(x|x')$ and $G_m(x|x')$ are the corresponding retarded
Green's functions which satisfy
\begin{eqnarray}\nonumber
\Box G(x|x')&=&-\frac{4\pi}{\sqrt{-g}}\delta^4(x-x'),\\\nonumber
(\Box -m^2) G_m(x|x')&=&-\frac{4\pi}{\sqrt{-g}}\delta^4(x-x') .
\end{eqnarray}
Both Green's functions vanish for all points $x$ which are outside
the future light cone of $x'$. We assume that the spacetime is
globally hyperbolic, and therefore $G$ and $G_m$ are uniquely
determined by the above requirements. We further assume that the
integrals in equations (\ref{phix},\ref{phixm}) converge in some
local neighborhood of the particle's world line, but not on the
world line itself. Expressing $\Delta\phi$ with these Green
functions gives
\begin{equation}\label{psi2}
\Delta\phi(x)=q\int_{-\infty}^{\infty}\Delta G(x|z(\tau))d\tau\,,
\end{equation}
where $\Delta G(x|x')\equiv G(x|x')-G_m(x|x')$. Consider the above
expression for ${\Delta\phi}(x)$ in some local neighborhood of the
particle's world line. In the discussion below we will use two
separate expressions for $\Delta G(x|z)$ in Eq. (\ref{psi2}): a
local expression -- for the case where $z$ is in a local
neighborhood (defined below) of $x$, and a non-local expression
for the case where $z$ is outside this local neighborhood. In the
next subsection we derive the local expression for $\Delta G$.

%%%%%%%%%%%%%%%%%%%%%%%%%%%%%%%%%%%%%%%%%%%%%%%%%%%%%%%%%%%%%%%%%%%%%%%%%%%%%%%%%%%%%%%%%%%%%%%%%%%%%%%%%
\subsection{Local expression for $\Delta G$}

In the vicinity of any point $x'$ there is a local neighborhood,
 in which any two points can be connected by a unique
geodesic within this neighborhood, see theorem 1.2.2 in
\cite{Friedlander} (this neighborhood is sometimes called a
geodesically convex domain). In this neighborhood, the massless
and massive retarded Green's functions are given by \cite{Gdec}
\begin{eqnarray}\label{gdecomp}
G(x|x')&=&\Theta(\Sigma(x),x')[U(x|x')\delta(\sigma)-V(x|x')\Theta(-\sigma)],\\\label{gmdecomp}
G_m(x|x')&=&\Theta(\Sigma(x),x')[U(x|x')\delta(\sigma)-V_m(x|x')\Theta(-\sigma)].
\end{eqnarray}
Here, $\sigma=\sigma(x|x')$ is half the square of the invariant
distance measured along a geodesic connecting $x$ and $x'$;
$\sigma$ is negative for a timelike geodesic, positive for a
spacelike geodesic, and vanishes for a null geodesic; $U(x|x')$,
$V(x|x')$, and $V_m(x|x')$ are certain bi-scalars (for their
definitions and properties, see \cite{Friedlander}); $\Sigma(x)$ is
an arbitrary space-like hypersurface containing $x$; and
$\Theta(\Sigma(x),x')$ equals unity if $x'$ is in the past of
$\Sigma(x)$ and vanishes otherwise.

From equations (\ref{gdecomp},\ref{gmdecomp}) we find that locally
(i.e., in a geodesically convex domain)
\begin{equation}\label{delg}
\Delta G(x|x')=-\Delta
V(x|x')\Theta(\Sigma(x),x')\Theta(-\sigma)\,,
\end{equation}
where,
\begin{equation}\label{defdelv}
\Delta V(x|x')\equiv{V}(x|x')-V_m(x|x')\, .
\end{equation}
Note that the {\em{same}} "direct term" $U(x|x')\delta(\sigma)$
appears in both Green's functions in equations (\ref{gdecomp}) and
(\ref{gmdecomp}), and therefore it cancels upon their subtraction.

Each of the direct fields
 $\phi_{dir}$ and $\phi_{m(dir)}$, which were mentioned in the
 previous section, is obtained
 by integrating over the corresponding direct term. These fields are
 therefore the same, and are given by
\begin{equation}\label{direql}
\phi_{dir}(x)=\phi_{m(dir)}(x)=q\int_{\tau^--\varepsilon}^{\infty}\Theta(\Sigma(x),z)U(x|z)\delta(\sigma)d\tau\,.
\end{equation}
 Here $\tau^-$ denotes the retarded proper time [i.e., the
proper time at the point of intersection of the past null cone of
the field evaluation point $x$, with the particle's world line
$z(\tau)$], and $\varepsilon$ is an arbitrary small time interval.
Since these direct fields are equal, they cancel each other in Eq.
(\ref{deltaphi}), which gives Eq. (\ref{del2}).

It will be useful later to have the explicit dependence of $V_m$
on $m$; for this purpose we use the Hadamard expansion.
 It was shown by Hadamard
\cite{Hadamard} that the bi-scalars $V$ and $V_m$ can be expanded
as follows\footnote{These two expansions are defined slightly
different from the various definitions in \cite{DB},
\cite{Hadamard}, and \cite {Friedlander}. The coefficients of
these various definitions are related to the coefficients in
equations (\ref{vexp},\ref{vmexp}) by multiplying by certain
numbers.}:
\begin{eqnarray}\label{vexp}
&&V(x|x')=\sum_{n=0}^{\infty}\frac{\sigma^n}{n!}v_n(x|x')\,,\\
&&V_m(x|x')=\sum_{n=0}^{\infty}\frac{\sigma^n}{n!}\tilde{v}_n(x|x')\,,\label{vmexp}
\end{eqnarray}
where the coefficients $v_n$ and $\tilde{v}_n$ satisfy certain
differential equations.
 In the case of an analytic metric, for every point $x'$ there is
  a local neighborhood in which the Hadamard expansion is uniformly
convergent \cite{Hadamard} (This is also true for a class of
non-analytic metrics \cite{Riesz}.). The local neighborhood of
$x'$ in which both equations (\ref{vexp},\ref{vmexp}) converge
uniformly is denoted by $D(x')$. Here and below we shall consider
the Hadamard expansion inside $D(x')$. Expressing the coefficients
$\tilde{v}_n$ in terms of the coefficients $v_n$, and the
bi-scalar $U$ gives [The derivation of this expression is given in
Appendix A, for a different derivation, see Friedlander
\cite{Friedlander} equation (6.4.19).]
\begin{equation}\label{vmbessel}
V_m(x|x')= mU\frac{J_1(ms)}{s}+\sum_{n=0}^{\infty}v_n
J_n(ms)\left(\frac{-s}{m}\right)^n\,.
\end{equation}
Here $J_n$ denotes the Bessel function, and we introduced
$s\equiv\sqrt{-2\sigma}$. Note that the bi-scalar $V_m$ is an even
function of $s$. We point out that the local expression for
$\Delta G$ inside $D(x')$ is obtained by substituting equations
(\ref{vexp},\ref{vmbessel}) in equations
(\ref{delg},\ref{defdelv}).

Before continuing the detailed calculation, we briefly discuss
some of the properties of $V_m$, and their relations to the self
force in our method. In the prescription given by Eq. (\ref{sf})
the asymptotic expression of
$\lim_{\delta\rightarrow0}{\Delta\phi}_{,\mu}$ , as
$m\rightarrow\infty$ is required. Now, Eq. (\ref{psi2}) implies
that $\Delta\phi$ can be determined by an integral over $\Delta
G$, and by equations (\ref{delg},\ref{defdelv}) the local
expression of $\Delta G$ depends  on $V$ and $V_m$. Therefore, to
obtain the asymptotic form (as $m\rightarrow\infty$) in Eq.
(\ref{sf}), we have to study the asymptotic form of $V_m$ (as well
as the asymptotic form of the nonlocal expression of $\Delta G$).
From Eq. (\ref{vmbessel}) we find that
 as $m\rightarrow\infty$ the Bessel functions oscillate rapidly (for $s\ne0$);
these rapid oscillations  have a canceling effect upon
integration with respect to $s$. In fact, the calculations below
show that at the limit $m\rightarrow\infty$ an integral (with
respect to $s$) over the first term in Eq. (\ref{vmbessel}) behaves
much like an integral over the direct term $U\delta(\sigma)$.
 This is an important property
in our construction\footnote{A similar property was used by
Coleman in the calculation in flat spacetime \cite{Coleman}.},
because it explains the source of the local terms of the self
force in our method. In other methods, for example in
\cite{Quinn}, the local terms of the self-force are essentially
obtained from the value of ${\phi_{dir}}_{,\mu}$ near the world
line, by some regularization method. Here however, in the
expression for $\Delta\phi$ the direct fields  ${\phi_{dir}}$ and
${\phi_{m(dir)}}$ canceled upon subtraction. Therefore it might
seem strange that we still obtain the local terms of the
self-force.
 This is possible because at the limit
$m\rightarrow\infty$ we find that the first term in Eq.
(\ref{vmbessel}) behaves similarly to the above mentioned direct term.

%%%%%%%%%%%%%%%%%%%%%%%%%%%%%%%%%%%%%%%%%%%%%%%%%%%%%%%%%%%%%%%%%%%%%%%%%%%%%%%%%%%%%%%%%%%%%%%%%%%%%%%%%
\subsection{Calculation of ${\Delta F}_\mu$ near the particle's world line}

We now use the above local expression for $\Delta G$, and derive
an expression for the field ${\Delta F}_\mu(x)$ near the world
line $z(\tau)$. Before proceeding, we define some notation. The
particle's proper time at the self force evaluation point $z_0$ is
denoted by $\tau_0$; The past intersection of the boundary of
$D(z_0)$ with the world line is denoted by $z(\tau_1)$. By our
construction the points $z_0$ and any point $z(\tilde{\tau})$,
such that $\tau_0>\tilde{\tau}\ge\tau_1$, can be connected by a
unique geodesic within $D(z_0)$. It may be that all these
geodesics are timelike geodesics . If this requirement is not
initially satisfied, we then increase the value of $\tau_1$
 such that all these geodesics with $\tau_0>\tilde{\tau}\ge\tau_1$ will become
timelike geodesics. The field evaluation point $x$ is taken to be
sufficiently close to the particle's world line (i.e. $\delta$ is
sufficiently small), such that both
 points $x$ and $z(\tau^-)$ are within $D(z_0)$ and $\tau^->\tau_1$.

Using equations (\ref{psi2},\ref{delg}), we obtain
\begin{equation}\label{delphint}
\Delta\phi(x)= -q\int_{\tau_1}^{\tau^-}\Delta V(x|z)\, d\tau
+q\int_{-\infty}^{\tau_1}\Delta G(x|z)\, d\tau.
\end{equation}
This field has a finite value as $x$ approaches the world line
(see Appendix \ref{regulardphi}). From equations
(\ref{deltafmu},\ref{delphint}) we find that
\begin{equation}\label{locnloc}
{\Delta F}_{\mu}(x)= -q^2 {(\tau^-)_{,}}_{\mu} \left[\Delta
V\right]_{\tau^-}- q^2\int_{\tau_1}^{\tau^-}\Delta
{V_,}_{\mu}d\tau +q^2\int_{-\infty}^{\tau_1}\Delta
{G_,}_{\mu}d\tau\, .
\end{equation}
Throughout this manuscript the indices $\mu,\nu$ refer to the
field evaluation point (here this point is denoted by $x$), and
the subscript $\tau^-$ indicates that the variable $z(\tau)$
inside the brackets is evaluated at the retardation point
$z(\tau^-)$. Using equations
(\ref{defdelv},\ref{vexp},\ref{vmbessel}) we obtain
\[
 \left[\Delta
V(x|z)\right]_{\tau^-}=-\frac{1}{2}m^2[U(x|z)]_{\tau^-}\,.
\]
This relation together with Eq. (\ref{locnloc}) give the following
expression for ${\Delta F}_\mu$
\begin{equation}\label{gradpsi0}
{\Delta F}_{\mu}(x)
={F^{tail}}_{\mu}(x)+\frac{1}{2}q^2m^2[U]_{\tau^-}{(\tau^-)_{,}}_{\mu}+
q^2\int_{\tau_1}^{\tau^-}{V_m}_{,\mu}d\tau
-q^2\int_{-\infty}^{\tau_1}{G_{m,}}_{\mu}d\tau\,.
\end{equation}
Here we introduced,
\begin{equation}\label{tailf}
{F^{tail}}_{\mu}(x)\equiv q^2 \int_{-\infty}^{\tau_1}
{G_{,}}_{\mu}d\tau-q^2\int_{\tau_1}^{\tau^-}{V_{,}}_{\mu}d\tau\,.
\end{equation}

It is useful for the calculations below, to change the splitting
of the integration interval in Eq. (\ref{gradpsi0}) into a "smooth
splitting" defined as follows. We introduce auxiliary weight
functions $h_1(\tau)$ and $h_2(\tau)$ (sufficiently smooth, see
below), and an arbitrarily small time interval $\epsilon$, such
that $h_1(\tau)\equiv1$ for $\tau\ge\tau_1+\epsilon$,
$h_1(\tau)\equiv0$ for $\tau\le\tau_1$, and $h_1(\tau)$ varies
smoothly in the interval $[\tau_1,\tau_1+\epsilon]$. Defining
 $h_2(\tau)\equiv1-h_1(\tau)$, we find from
Eq. (\ref{gradpsi0}) that
\begin{equation}\label{gradpsi} {\Delta
F}_{\mu}(x)
={F^{tail}}_{\mu}(x)+\frac{1}{2}q^2m^2[U]_{\tau^-}{(\tau^-)_{,}}_{\mu}+
q^2\int_{\tau_1}^{\tau^-}h_1{V_m}_{,\mu}d\tau-
q^2\int_{-\infty}^{\tau_1+\epsilon}h_2{G_{m,}}_{\mu}d\tau\,.
\end{equation}
Eq. (\ref{gradpsi}) provides a general formal expression for the
field ${\Delta F}_{\mu}$ near the object's world line. Next, we
shall use Eq. (\ref{gradpsi}) to show that Eq. (\ref{sf}) is
identical to the scalar self force.
%%%%%%%%%%%%%%%%%%%%%%%%%%%%%%%%%%%%%%%%%%%%%%%%%%%%%%%%%%%%%%%%%%%%%%%%%%%%%%%%%%%%%%%%%%%%%%%%%%%%%%%%%
\subsection{Derivation of the self force expression }

We now follow the prescription given by Eq. (\ref{sf}), and
perform the following successive operations on the field ${\Delta
F}_{\mu}$:
\newcounter{bean}
\begin{list}{\roman{bean}}{\usecounter{bean}}
  \item Calculating the limit $\delta \rightarrow 0$ of ${\Delta F}_{\mu}$.
  \item Calculating the asymptotic form of $\lim_{\delta \rightarrow 0}{\Delta F}_{\mu}$
   as $m$ approaches infinity.
\end{list}
These mathematical operations will be performed separately on each
term in equation (\ref{gradpsi}).
%%%%%%%%%%%%%%%%%%%%%%%%%%%%%%%%%%%%%%% first term %%%%%%%%%%%%%%%%%%%%%%%%%%%%%%%%%%%%%
\subsubsection{The first term}

Consider first the limit ${\delta\rightarrow0}$ of the first term
${F^{tail}}_{\mu}$ in Eq. (\ref{gradpsi}). Since $V$ is smooth
(see theorem 4.5.1 in \cite{Friedlander}), and
 $\tau^-(x)\rightarrow\tau_0$ as $\delta\rightarrow 0$,
 we find from Eq. (\ref{tailf}) that
 \begin{equation}\label{fstterm}
 \lim_{\delta\rightarrow
 0}{F^{tail}}_{\mu}(x)=q^2 \int_{-\infty}^{\tau_1}
{G_{,}}_{\mu}(z_0|z(\tau))d\tau-q^2\int_{\tau_1}^{\tau_0}{V_{,}}_{\mu}
(z_0|z(\tau))d\tau\, .
\end{equation}
 This expression is equivalent
 to the non-local term of the scalar self-force which was found in Ref. \cite{Quinn}
. Since, this term is independent of $m$, it is not affected by
limit $m\rightarrow\infty$.

%%%%%%%%%%%%%%%%%%%%%%%%%%%%%%%%%%%%%%%%%%%%%% second term %%%%%%%%%%%%%%%%%%%%%%%%%%%%%%%%%%%

\subsubsection{The second term}

Consider next the second term in  Eq. (\ref{gradpsi}). The second
term can be expressed as \footnote{This expression follows from
the
 relation between $\tau^-$ and $x$ on the lightcone,
  which is given by the equation $\sigma(x|z(\tau^-))=0$; see for
example \cite{Quinn}.}
\begin{equation}\label{gradtaum}
\frac{1}{2} q^2m^2(U)_{\tau^-}{(\tau^-)}_{,\mu}=
-\frac{1}{2}q^2m^2(U)_{\tau^-}\left(\frac{{\sigma}_{,\mu}}{\sigma_{,\alpha}u^\alpha}
\right)_{\tau^-}\,.
\end{equation}
Here the index $\alpha$ refers to the point $z^\alpha(\tau)$ on
the particle's world line, and $u^\alpha\equiv
\frac{dz^\alpha}{d\tau}$. The calculation of the limit
$\delta\rightarrow0$ of this expression follows from a local
expansion of the various terms in Eq. (\ref{gradtaum}). The
detailed calculation is given in Appendix B, there we show that
\begin{equation}\label{tauret}
\lim_{\delta\rightarrow0}\frac{1}{2}q^2m^2(U)_{\tau^-}{(\tau^-)_{,}}_{\mu}=
-\frac{1}{2}q^2m^2\left(u_\mu+n_\mu\right)\,.
\end{equation}
%%%%%%%%%%%%%%%%%%%%%%%%%%%%%%%%%%%%%%%%%% third term %%%%%%%%%%%%%%%%%%%%%%%%%%%%%%%%%%%%%%%%%%%%
\subsubsection{The third term}

Consider next the third term in Eq. (\ref{gradpsi}). First, we
take the limit ${\delta\rightarrow0}$ of this term. Similar to
 Eq. (\ref{fstterm}) we find that
\[
\lim_{\delta\rightarrow
0}\int_{\tau_1}^{\tau^-}h_1{V_m}_{,\mu}(x|z(\tau))d\tau=
\int_{\tau_1}^{\tau_0}h_1{V_m}_{,\mu}(z_0|z(\tau))d\tau\, .
\]
We now split the integration interval into two intervals: one
interval is $[\tau_1+\epsilon,\tau_0]$ in which $h_1\equiv1$, and
the the other interval is $[\tau_1,\tau_1+\epsilon]$ in which
$h_1$ varies smoothly, this gives
\begin{equation}\label{limint}
\int_{\tau_1}^{\tau_0}h_1{V_m}_{,\mu}(z_0|z(\tau))d\tau=
\int_{\tau_1+\epsilon}^{\tau_0}{V_m}_{,\mu}(z_0|z(\tau))d\tau
+\int_{\tau_1}^{\tau_1+\epsilon}h_1{V_m}_{,\mu}(z_0|z(\tau))d\tau\,.
\end{equation}
First we focus on the first integral on the right hand side of Eq.
(\ref{limint}). We change the integration variable to $s$ (with
$z_0$ fixed). Recall that for any point $z(\tilde{\tau})$ on the
world line such that  $\tau_1\le \tilde{\tau}<\tau_0$, we have a
unique timelike geodesic within $D(z_0)$, which connects $z_0$ and
$z(\tilde{\tau})$. For each $z(\tilde{\tau})$ in this range we
have
\[
\frac{ds}{d\tau}(z_0|z(\tilde{\tau}))=s_{,\alpha}\frac{dz^\alpha}{d\tau}<0\,.\]
This expression is negative because $s_{,\alpha}$ is a timelike
(future directed) vector at $z(\tilde{\tau})$ which is tangent to
the geodesic that connects $z(\tilde{\tau})$ and $z_0$; and
$\frac{dz^\alpha}{d\tau}$ is the four velocity vector (also future
directed) at $z(\tilde{\tau})$.

 Substituting Eq.
(\ref{vmbessel}) into the first integral in Eq. (\ref{limint}),
and interchanging the order of integration and summation gives
\begin{eqnarray}\nonumber
&&\int_{\tau_1+\epsilon}^{\tau_0}{V_m}_{,\mu}(z_0|z(\tau))d\tau=
\\\nonumber&&\sum_{n=-1}^{\infty}
{(-m)}^{-n}\int_{\tilde{s}_1}^0\bigglb[{{v_n}_{,}}_{\mu}J_n(ms){s}^n+
 v_n
\frac{d}{ds}
\left[J_n(ms){s}^n\right]{s_{,}}_{\mu}\biggrb]\frac{d\tau}{ds}ds \,
.
\end{eqnarray}
Here we introduced $\tilde{s}_1\equiv s(z_0|z(\tau_1+\epsilon))$,
$v_{-1}\equiv U$, and the various bi-quantities are evaluated at
the points $z_0$ and $z(\tau)$ [e.g., $s=s(z_0|z(\tau))$,
$v_n=v_n(z_0|z(\tau))$ etc ...], this notation is used in the rest
of this subsection. Using integration by parts we find that
\begin{eqnarray}\label{localp}
&&\int_{\tau_1+\epsilon}^{\tau_0}{V_m}_{,\mu}(z_0|z(\tau))d\tau=\\\nonumber
&&\sum_{n=-1}^{\infty}\int_{\tilde{s}_1}^0J_n(ms)\frac{(-s)^n}{m^n}
\left[\frac{d\tau}{ds}\left({v_{n,}}_{\mu}-{s_,}_{\mu}\frac{\partial
v_n}{\partial s}\right) -\frac{\partial}{\partial
s}\left(\frac{d\tau}{ds}{s_,}_{\mu}\right)v_{n}\right]
 ds+\\\nonumber
 &&\left({V_m}\frac{d\tau}{ds}{s_,}_{\mu}\right)_{s=0}-
 \left({V_m}\frac{d\tau}{ds}{s_,}_{\mu}\right)_{s=\tilde{s}_1}\, .
\end{eqnarray}
Here, in the differentiation with respect to $s$, $z_0$ is fixed;
whereas in the differentiation with respect to $x^\mu$,
$z^\alpha(\tau)$ is fixed. Local analysis shows that the first
boundary term in Eq. (\ref{localp}) is given by (see Appendix B)
\begin{equation}\label{boundt1}
\left({V_m}\frac{d\tau}{ds}{s_,}_{\mu}\right)_{s=0}
=\left(\frac{m^2}{2}-\frac{R}{12}\right)u_\mu\,.
\end{equation}
Here $R$ denotes the Ricci scalar\footnote{Here the Riemann tensor is
defined with the opposite sign with respect to the definition in
\cite{DB}.}. We do not need to calculate the second boundary term
in Eq. (\ref{localp}) since it cancels with another term (see
below).

Next, we calculate the asymptotic form of the integral in Eq.
(\ref{localp}) as $m\rightarrow\infty$. Consider the integral over
the n'th term in this equation. To calculate the asymptotic form
of this integral, we expand the term inside the square brackets in
powers of $s$. Integration over the terms in this expansion
requires one to calculate integrals of the following form:
\begin{equation}\label{intj}
\int_{\tilde{s}_1}^0J_n(ms)\frac{s^{n+\alpha}}{m^n}ds
=\frac{1}{m^{2n+\alpha+1}}\int_{m\tilde{s}_1}^0J_n(y)y^{n+\alpha}
dy\,.
\end{equation}
Note that the coefficients $v_n$ in Eq. (\ref{localp}) are all
smooth functions (see  \cite{Friedlander}, sec. 4.3), and the
expansion of the expressions that do not depend on $v_n$, contain
only positive powers of $s$ (see Appendix B). Therefore, in Eq.
(\ref{intj}) we need to consider only $\alpha\ge0$. As
$m\rightarrow\infty$ the integral in this equation vanishes for
$n\ge0$. Therefore, only the terms with $n=-1$ contribute to Eq.
(\ref{localp}). In this case ($n=-1$) the expansion of the term in
the square brackets in Eq. (\ref{localp}) includes only integer
powers of $s$ (see below). From these terms only the terms with
$\alpha=0,1$ do not vanish as $m\rightarrow\infty$. Therefore, we
need to expand the $n=-1$ terms only up to the first order in
$s$. These expansions give (see Appendix B)
\begin{eqnarray}\label{ex1s}
&&U\frac{\partial}{\partial s
}\left(s_{,\mu}\frac{d\tau}{ds}\right)=
-\frac{1}{2}a_\mu+\frac{1}{3}(\dot{a}_\mu-a^2u_\mu)s+O(s^2)\,,\\
\label{ex2s}
&&\frac{d\tau}{ds}\left({U_{,}}_{\mu}-{s_,}_{\mu}\frac{\partial
U}{\partial s}\right)= -\frac{1}{6}\left({R_\mu}^\nu u_\nu+
R^{\eta\nu} u_\eta u_\nu u_\mu \right)s+O(s^2)\, .
\end{eqnarray}
Here $\dot{a}_\mu$ denotes the covariant derivative of $a_\mu$
with respect to $\tau$. Here and below (unless explicitly
indicated otherwise) the coefficients
$u_\mu,a_\mu,\dot{a}_\mu,R^{\mu\nu},R $ are evaluated at $z_0$.
Equations
(\ref{localp},\ref{boundt1},\ref{intj},\ref{ex1s},\ref{ex2s}) give
\begin{eqnarray}\label{ferterm}
&&\int_{\tau_1+\epsilon}^{\tau_0}{{V_m}_{,\mu}(z_0|z(\tau))}d\tau\cong\\\nonumber
&&\frac{1}{2}m^2u_\mu-\frac{1}{2}ma_\mu+\frac{1}{3}(\dot{a}_\mu-a^2u_\mu)+
\left(\frac{1}{6}{R_\mu}^\nu u_\nu+ \frac{1}{6}R^{\eta\nu} u_\eta
u_\nu u_\mu
-\frac{1}{12}R{u}_\mu\right)\\\nonumber
&&- \left({V_m}\frac{d\tau}{ds}{s_,}_{\mu}\right)_{s=\tilde{s}_1}\,.
\end{eqnarray}
Here, and throughout this manuscript, the symbol $\cong$
represents equality up to terms that vanish as $m\rightarrow
\infty$.

We now focus on the second integral on the right hand side of Eq.
(\ref{limint}). Using a calculation similar to the one that was
performed on the first integral in  Eq. (\ref{limint}), we find
that
\begin{eqnarray}\label{secterm}
&&\int_{\tau_1}^{\tau_1+\epsilon}h_1{V_m}_{,\mu}(z_0|z(\tau))d\tau=\\\nonumber
&&\sum_{n=-1}^{\infty}\int_{s_1}^{\tilde{s}_1}J_n(ms)\frac{(-s)^n}{m^n}
\left[\frac{d\tau}{ds}h_1\left({v_{n,}}_{\mu}-{s_,}_{\mu}\frac{\partial
v_n}{\partial s}\right) -\frac{\partial}{\partial
s}\left(h_1\frac{d\tau}{ds}{s_,}_{\mu}\right)v_{n}\right]
 ds+\\\nonumber
 &&\left(h_1{V_m}\frac{d\tau}{ds}{s_,}_{\mu}\right)_{s=\tilde{s}_1}-
 \left({V_m}h_1\frac{d\tau}{ds}{s_,}_{\mu}\right)_{s=s_1}\, .
\end{eqnarray}
Here we introduced ${s}_1\equiv s(z_0|z(\tau_1))$. Recall that
$h_1(s_1)=0$, and $h_1(\tilde{s}_1)=1$. Therefore, the second
boundary term in Eq. (\ref{secterm}) vanishes, and the first
boundary term is minus the boundary term in Eq. (\ref{ferterm}).
Noting that asymptotically $J_n(x)=\sqrt{\frac{2}{\pi
x}}\cos[x-(n+\frac{1}{2})\frac{\pi}{2}]+O(x^{-3/2})$ \cite
{Arfken}, we find that as $m\rightarrow \infty$ the integral in
Eq. (\ref{secterm}) vanishes for $n\ge0$. For $n=-1$ we substitute
$J_{-1}(ms)=\frac{1}{m}\frac{d}{ds}J_0(ms)$ in this integral, and
integrate by parts once more. Taking the limit
$m\rightarrow\infty$ we find that\footnote {Here we require that
$h_1(s)$ is a $C^2$ function.}
\begin{equation}\label{secte2}
\int_{\tau_1}^{\tau_1+\epsilon}h_1{{V_m}_{,\mu}(z_0|z(\tau))}d\tau\cong
 \left({V_m}\frac{d\tau}{ds}{s_,}_{\mu}\right)_{s=\tilde{s}_1}\,.
\end{equation}
From equations (\ref{limint},\ref{ferterm},\ref{secte2}) we obtain
\begin{eqnarray}\label{thirdf}
&&q^2\int_{\tau_1}^{\tau_0}h_1{{V_m}_{,\mu}(z_0|z(\tau))}d\tau\cong\\\nonumber
&&q^2\bigglb\{\frac{1}{2}m^2u_\mu-\frac{1}{2}ma_\mu+
\frac{1}{3}(\dot{a}_\mu-a^2u_\mu)\\\nonumber
&&+\biglb(\frac{1}{6}{R_\mu}^\nu u_\nu+ \frac{1}{6}R^{\eta\nu} u_\eta
u_\nu u_\mu
 -\frac{1}{12}R{u}_\mu\bigrb)\biggrb\}\,.
\end{eqnarray}
Note that the local terms of the scalar self force found in Ref.
\cite{Quinn} are identical to the terms that are independent of
$m$ in Eq. (\ref{thirdf}).

%%%%%%%%%%%%%%%%%%%%%%%%%%%%%%%%%%%%%%%%%%% fourth term %%%%%%%%%%%%%%%%%%%%%%%%%%%%%%%%%%%%%%%%%%%%%%
\subsubsection{The fourth term}

Consider next the fourth term on the right hand side in Eq.\
(\ref{gradpsi}). Taking the limit $\delta\rightarrow0$ of this
term gives
\begin{equation}\label{fterm}
\int_{-\infty}^{\tau_1+\epsilon}h_2{G_{m,}}_{\mu}(z_0|z(\tau))d\tau\,
.
\end{equation}
We now discuss the limit $m\rightarrow \infty$ of this expression.
Notice that in the above calculation of the third term, we found
that the value of the third term at the limit $m\rightarrow
\infty$ does not depend on $\tau_1$ [see Eq. (\ref{thirdf})]. This
suggests that the entire fourth term should vanish at the limit of
interest. To further support this statement we introduce the
following physically motivated assumption. It is well known that
in flat spacetime as $m\rightarrow \infty$ the range of the
massive field interaction approaches zero - here we shall assume
that this statement is also valid in curved spacetime. Therefore,
the massive scalar force $F_{(m)\mu}$ for any point that is not
on the world line vanishes as $m\rightarrow\infty$. We now
consider a different world line $\tilde{z}(\tilde{\tau})$, which
coincides with $z(\tau)$ for $\tau\le\tau_1+\epsilon$, and is
slightly displaced with respect to $z(\tau)$ for
$\tau>\tau_1+\epsilon$, such that $z_0$ is {\em not} on
$\tilde{z}(\tilde{\tau})$. The massive force field
$F_{(m)\mu}(z_0)$ for the world line $\tilde{z}$ is given by
\[
F_{(m)\mu}(z_0)=q^2\partial_\mu\int_{-\infty}^{{\tilde{\tau}}^{-}}{G_{m}}(z_0|\tilde{z}(\tilde{\tau}))
d\tilde{\tau}\,
 \]
 where ${{\tilde{\tau}}^{-}}$ is the retarded proper time on
$\tilde{z}$.
 Using the same smooth splitting
as in Eq. (\ref{gradpsi}), we find that
\begin{equation}\label{fmmuz0}
F_{(m)\mu}(z_0)=q^2\int_{-\infty}^{\tau_1+\epsilon}h_2{G_{m,}}_{\mu}d\tau+
 q^2\partial_\mu\int_{\tau_1}^{{\tilde{\tau}}^{-}}h_1{G_{m}}d\tilde{\tau}\,.
 \end{equation}
 Notice that the first term in this equation is identical to Eq. (\ref{fterm}).
By calculating the limit $m\rightarrow \infty$ of the second
integral in Eq. (\ref{fmmuz0}) one finds that this term vanishes
at this limit \footnote {This calculation is based on the Hadamard
expansion, and is similar to the calculation of the third term
given above.}. Therefore,  Eq. (\ref{fmmuz0}) together with our
above assumption imply that the entire forth term vanishes at the
limit of interest. We verify this conclusion for the special case
of a de Sitter background spacetime (see Appendix D).

%%%%%%%%%%%%%%%%%%%%%%%%%%%%%%%%%%%%%%%%%%%%% result  %%%%%%%%%%%%%%%%%%%%%%%%%%%%%%%%%%%%%%%%%%%%%%%%%
\subsubsection{Result}

By substituting the above four terms in equation
(\ref{gradpsi}) we obtain
\begin{eqnarray}\label{psimuapp}
&&\lim_{\delta \rightarrow 0
}{\Delta F}_{\mu}\cong{F^{tail}}_{\mu}(z_0)+\\\nonumber
&&q^2\bigglb\{-\frac{1}{2}m^2n_\mu-\frac{1}{2}ma_\mu+\frac{1}{3}(\dot{a}_\mu-a^2u_\mu)\\\nonumber
&&+\bigglb(\frac{1}{6}{R_\mu}^\nu u_\nu+ \frac{1}{6}R^{\eta\nu} u_\eta
u_\nu u_\mu -\frac{1}{12}Ru_\mu\biggrb)\biggrb\}\,.
\end{eqnarray}

The term in the curly brackets in Eq. (\ref{psimuapp}) has a
directional dependence on $n_\mu$ and it diverges as $m\rightarrow
\infty$. To remove this directional dependence, and the divergence
behavior as $m\rightarrow \infty$; we subtract the term
$q^2(-\frac{1}{2}m^2n_\mu-\frac{1}{2}ma_\mu)$ from both sides of
this equation. We can now safely take the limit $m \rightarrow
 \infty$, from which we obtain Eq. (\ref{sf})
\begin{eqnarray}
&&f_\mu^{self}(z_0)=q\lim_{m\rightarrow\infty}\left\{
\lim_{\delta\rightarrow 0} \Delta\phi_{,\mu}+
\frac{1}{2}qm^2n_\mu+\frac{1}{2}qma_\mu\right\} =\\\nonumber
&&{F^{tail}}_{\mu}(z_0)+
q^2\left[\frac{1}{3}(\dot{a}_\mu-a^2u_\mu)+
\left(\frac{1}{6}{R_\mu}^\nu u_\nu+ \frac{1}{6}R^{\eta\nu} u_\eta
u_\nu u_\mu -\frac{1}{12}Ru_\mu\right)\right]\,.
\end{eqnarray}
This last expression is identical to the scalar self force
expression found in \cite{Quinn}.

%%%%%%%%%%%%%%%%%%%%%%%%%%%%%%%%  acknowledgments   %%%%%%%%%%%%%%%%%%%%%%%%%
\subsection*{Acknowledgments}
I am grateful to Amos Ori for valuable discussions. I wish to
thank the organizers of the sixth Capra meeting on radiation
reaction for the generous hospitality, and the participants
at this meeting, for discussions.

%%%%%%%%%%%%%%%%%%%%%%%%%%%%%%%%%%%% APPENDIX A%%%%%%%%%%%%%%%%%%%%%%%%%%%%%%%
\appendix\label{appa}
\section{Coefficients relations}

Here we express the coefficients $\tilde{v}_n(x|x')$ in terms of
the coefficients $v_n(x|x')$ and the bi-scalar $U(x|x')$. From
these relations we then derive Eq. (\ref{vmbessel}).

We consider $x$ to be within $D(x')$. The coefficients $v_n(x|x')$
and $\tilde{v}_n(x|x')$ satisfy the following recurrence
differential
  equations along the geodesics which connects $x$ and $x'$ (see \cite{Friedlander}, Sec.
  4.3):
\begin{eqnarray}\label{vneq}
&&\tilde{v}_0+\left(\tilde{v}_{0;\mu}-\frac{\tilde{v}_0}{U}U_{;\mu}\right)
 {\sigma_;}^{ \mu}=-\frac{1}{2}(\Box-m^2)U\,, \\\label{vneq2}
&&\tilde{v}_n(n+1)+\left(\tilde{v}_{n_;\mu}-\frac{\tilde{v}_n}{U}U_{;\mu}\right)
 {\sigma_;}^{ \mu}=-\frac{1}{2}(\Box-m^2)\tilde{v}_{n-1} \,.
\end{eqnarray}
Here the index $\mu$ refer to derivatives with respect to $x^\mu$.
The differential equations for the coefficients $v_n(x|x')$ are
obtained by substituting $m=0$ in equations
(\ref{vneq},\ref{vneq2}). We
 introduce an affine parameter $r$
along the geodesic $x(r)$ extending from $x'$ to $x$, where
$x(0)=x'$ and $x(1)=x$. By substituting
${\sigma_;}^{\mu}=r\frac{dx^\mu}{dr}$, into equations
(\ref{vneq},\ref{vneq2}), these equations can be integrated
\cite{Friedlander}, which gives
\begin{eqnarray}\label{vint0}
&&\tilde{v}_0(x(r)|x')=-\frac{U}{2r}\int_0^r\left[\frac{(\Box-m^2)U}{U}\right]_{x(r')}\,dr' \\
\label{vintn} &&\tilde{v}_n(x(r)|x')=-\frac{U}{2r^{n+1}}\int_0^r
{r'}^{n}
\left[\frac{(\Box-m^2)\tilde{v}_{n-1}}{U}\right]_{x(r')}\,dr'\,.
\end{eqnarray}
Similar expressions for $v_n(x(r)|x')$ are obtained by
substituting $m=0$ in these equations which gives
\begin{eqnarray}\label{v0m0}
&&{v}_0(x(r)|x')=-\frac{U}{2r}\int_0^r\left[\frac{\Box U}{U}\right]_{x(r')}\, dr' \\
\label{vnm0} &&{v}_n(x(r)|x')=-\frac{U}{2r^{n+1}}\int_0^r {
r'}^{n}\left[\frac{\Box {v}_{n-1}}{U}\right]_{x(r')}\,dr'\,.
\end{eqnarray}

First, we consider $\tilde{v}_0$. From equations
(\ref{vint0},\ref{v0m0}) we find that
\begin{equation}\label{v0}
\tilde{v}_0=v_0+\frac{1}{2}Um^2\,.
\end{equation}
Next we consider $\tilde{v}_1$. From equation (\ref{v0m0}) we
find that
\begin{equation}\label{partr}
r\frac{1/2\Box U-v_0}{U}= -\frac{\partial}{\partial
r}\left(\frac{v_0r^2}{U}\right)\,.
\end{equation}
By substituting  Eq. (\ref{v0}) into Eq. (\ref{vintn}) with $n=1$
and using Eq. (\ref{partr}), we obtain
\begin{equation}
\tilde{v}_1=v_1+\frac{v_0}{2}m^2+\frac{U}{8}m^4\,.
\end{equation}
By repeating this process of substitution of the coefficients
$\tilde{v}_i$ in  Eq. (\ref{vintn}) for $\tilde{v}_{i+1}$, we find
that $\tilde{v}_n$ is given by
\begin{equation}
\tilde{v}_n=
\sum_{k=-1}^n\left(\frac{m^2}{2}\right)^{n-k}\frac{v_k}{(n-k)!}\,,
\end{equation}
where $v_{-1}\equiv U$. The Hadamard expansion for $V_m$ in Eq.
(\ref{vmexp}) now takes the form
\begin{equation}\label{vmfirst}
V_m(x|x')= \sum_{n=0}^{\infty}\frac{\sigma^n}{n!}
\sum_{k=-1}^n\left(\frac{m^2}{2}\right)^{n-k}\frac{v_k}{(n-k)!}\,.
\end{equation}
We introduce $p=n-k$, and rearrange the order of the terms in Eq.
(\ref{vmfirst})\footnote{We do not discuss the question whether it
is permissible to rearrange the order of the terms in this sum. We
comment
 however that our resultant expression is equivalent to an
expression for $V_m$ obtained by Friedlander  in a completely
different manner, see \cite{Friedlander} equation (6.4.19).} which
gives
\begin{equation}
V_m(x|x')=\sum_{k=-1}^{\infty}\sigma^k v_k\sum_{p=0}^{\infty}
\frac{(-1)^p(ms/2)^{2p}}{p!(p+k)!}=
mU\frac{J_1(ms)}{s}+\sum_{k=0}^{\infty}v_k
J_k(ms)\left(\frac{-s}{m}\right)^k\,.
\end{equation}
Here $s\equiv\sqrt{-2\sigma}$.

%%%%%%%%%%%%%%%%%%%%%%%%%%%%%%%%%%%%%%%%%%%%%%% APPENDIX B%%%%%%%%%%%%%%%%%%%%%%%%%%%%%%%

\section{Local analysis}\label{appb}

Here we derive equations
(\ref{ex1s},\ref{ex2s},\ref{boundt1},\ref{tauret}). First,
consider the derivation of equation (\ref{ex1s}). This equation
requires a local expansion of the expression
$U\frac{d}{ds}(s_{,\mu}\frac{d\tau}{ds})$ in powers of $s$. We
start by expanding ${\sigma}_{;\mu}(z_0|z(\tau))$ for a point
$z(\tau)$ which is in the vicinity of point $z_0$, this gives
\begin{equation}
{\sigma}_{;\mu}(z_0|z(\tau_0-t))=
[{\sigma}_{;\mu}]-t[{\partial_\tau{\sigma}}_{;\mu}]+\frac{t^2}{2}[{\partial_\tau^2{\sigma}}_{;\mu}]-
\frac{t^3}{6}[{\partial_\tau^3{\sigma}}_{;\mu}]+O(t^4) .
\end{equation}
Here we introduced the proper time difference
$t\equiv\tau_0-\tau$, and the following notation for the
coincidence limit $
[{\sigma}_{;\mu}]\equiv\lim_{t\rightarrow0}{\sigma}_{;\mu}$. Here
and below the indices $\mu,\nu,\eta$ refer to the point
$x^\mu=z^\mu_0$, and the index $\alpha$ refers to the point
$z^\alpha(\tau_0-t)$. Employing the coincidence limits given in
\cite{DB,Synge} we obtain
\begin{equation}\label{gradsigma}
{\sigma}_{;\mu}(z_0|z(\tau))=
tu_\mu-\frac{t^2}{2}a_\mu+\frac{t^3}{6}\dot{a}_\mu+O(t^4)\,.
\end{equation}
Here and below, unless indicated otherwise the coefficients
$u_\mu,a_\mu,\dot{a}_\mu$ are evaluated at $z_0$. By
differentiating the normalization  $u^\mu(\tau) u_\mu(\tau)=-1$
twice, we find that $a^2=-\dot{a}^\mu u_\mu$. From this relation
together with $s^2=-\sigma_{;\mu}{\sigma_;}^{\mu}$ \footnote{For
this relation see \cite{DB}. Note, that here we defined
$s\equiv\sqrt{-2\sigma}$.} and  Eq. (\ref{gradsigma}) we obtain
\begin{equation}\label{sexp}
s(z_0|z(\tau))=t+\frac{1}{24}a^2t^3+O(t^4)\,.
\end{equation}
From this equation we find that
\begin{equation}\label{texpand}
t=s-\frac{1}{24}a^2s^3+O(s^4)\,,
\end{equation}
and therefore
\begin{equation}\label{dtauds}
\frac{d\tau}{ds}=-1+\frac{1}{8}a^2s^2+O(s^4)\,.
\end{equation}
From equations (\ref{gradsigma},\ref{sexp},\ref{texpand}) we
obtain
\begin{equation}\label{grads}
{s_,}_\mu(z_0|z(\tau))=-s^{-1}{\sigma_;}_\mu=-u_\mu+\frac{s}{2}a_\mu-
s^2\left(\frac{\dot{a}_\mu}{6}-\frac{u_\mu a^2}{24}\right)+O(s^3)\,.
\end{equation}
A local expansion of the bi-scalar $U$ \cite{DB} gives
\begin{equation}\label{uexp}
U=1+\frac{1}{12}R^{\mu\nu}(z_0)\sigma_{;\mu}\sigma_{;\nu}+O(s^3)\,.
\end{equation}
From equations (\ref{dtauds},\ref{grads},\ref{uexp}) we obtain Eq.
(\ref{ex1s})
\begin{equation}
U\frac{\partial}{\partial
s}(s_{,\mu}\frac{d\tau}{ds})=-\frac{1}{2}a_\mu+\frac{s}{3}(\dot{a}_\mu-a^2u_\mu)+O(s^2)\,.
\end{equation}

Next, consider the derivation of equation (\ref{ex2s}). Here our
purpose is to expand the expression $
\frac{d\tau}{ds}\left({U_{,}}_{\mu}-{s_,}_{\mu}\frac{\partial
U}{\partial s}\right)$ in powers of $s$. Recall that in the
differentiation with respect to $s$, $x^\mu$ is fixed; whereas in
the differentiation with respect to $x^\mu$, $z^\alpha(\tau)$ is
fixed. Using equations (\ref{dtauds},\ref{grads},\ref{uexp}), and
the above mentioned coincidence limits we obtain Eq. (\ref{ex2s})
\[
\frac{d\tau}{ds}\left({U_{,}}_{\mu}-{s_,}_{\mu}\frac{\partial
U}{\partial s}\right)= -\frac{1}{6}\left({R_\mu}^\nu u_\nu+
R^{\eta\nu} u_\eta u_\nu u_\mu \right)s+O(s^2)\,,
\]
where $R^{\eta\nu}$ is evaluated at $z_0$.
Next, consider the derivation of Eq. (\ref{boundt1}). From Eq.
(\ref{vmbessel}) we find that
\[
{(V_m)}_{s=0}=\frac{m^2}{2}{(U)}_{s=0}+{(v_0)}_{s=0}\,.
\]
Employing Eq. (\ref{uexp}), and noting that $(v_0)_{s=0}=-R/12$
\cite{DB}, we obtain
 \[
{(V_m)}_{s=0}=\frac{m^2}{2}-\frac{R}{12}\,,
\]
where $R$ is evaluated at $z_0$.
From this equation together with equations
(\ref{dtauds},\ref{grads}) we obtain Eq. (\ref{boundt1})
\[
 \left({V_m}\frac{d\tau}{ds}{s_,}_{\mu}\right)_{s=0}
=\left(\frac{m^2}{2}-\frac{R}{12}\right)u_\mu\,.
\]

Next, consider the derivation of equation (\ref{tauret}).
 This equation requires the calculation of the limit
 ${\delta\rightarrow0}$ of the expression
  ${(\tau^-)_{,}}_{\mu}=
  -\left[
  {\sigma}_{;\mu}{(\sigma_{;\alpha}u^\alpha)}^{-1}\right]_{\tau^-}$.
   We therefore expand the expression inside the
 brackets in powers of $\delta$, keeping only the first leading order
in this expansion. Consider first the term
${{\sigma}_{;\mu}}(x|z(\tau))$.  We choose the points
$x(\tau_0,n_\mu,\delta)$ and $z(\tau)$ to be in the vicinity of
the point $z(\tau_0)$, such that $\delta$ is of the same order as
$t$. Expanding ${{\sigma}_{;\mu}}$ we find that
\begin{equation}\label{dandtexp}
{\sigma}_{;\mu} (x(\tau_0,n_\mu,\delta)|z(\tau_0-t))=-
t\left[{\frac{dz^\alpha}{d\tau}{\sigma_{;\mu\alpha}}}\right]+
\delta\left[\frac{dx^\nu}{d\delta} \sigma_{;{\mu\nu}}\right]+O(\delta^2) \,,
\end{equation}
where the square brackets denote the coincidence limit
\[
[{\sigma}_{;\mu\nu}]\equiv\lim_{t\rightarrow0}\lim_{\delta\rightarrow0}{\sigma}_{;\mu\nu}\,.\]
Taking $\tau$ to be the retarded proper time $\tau=\tau^-$ gives
(see \cite{DB})
\begin{equation}\label{sigdot}
t=\delta+O(\delta^2)\,\,,\,\,
({\sigma_{;\alpha}u^\alpha})_{\tau^-}=\delta+O(\delta^2)\,.
\end{equation}
From equations
(\ref{gradtaum},\ref{uexp},\ref{dandtexp},\ref{sigdot}) together
with the above mentioned coincidence limits we obtain Eq.
(\ref{tauret})
\[\lim_{\delta\rightarrow0}\frac{1}{2}q^2m^2[U]_{\tau^-}{(\tau^-)_{,}}_{\mu}=
-\frac{1}{2}q^2m^2\left(u_\mu+n_\mu\right)\,.
\]
%%%%%%%%%%%%%%%%%%%%%%%%%%%%%%%%%%%%%%%%%%%%%% APPENDIX C %%%%%%%%%%%%%%%%%%%%%%%%%%%%%%%%%%%%%%%%%

\section{Regularity of $\Delta\phi(x)$ as $\delta\rightarrow0$} \label{regulardphi}

Here we show (under certain assumptions; see below)
 that $\Delta\phi(x)$ remains finite as $\delta\rightarrow0$.
We consider a finite value of $m$; hence the bi-scalar $\Delta
V=V-V_m$ is a smooth function (see theorem 4.5.1 in
\cite{Friedlander}). Therefore, the first integral in Eq.
(\ref{delphint}) is a continuous function of $\delta$ at $z_0$.

Consider next, the limit $\delta\rightarrow0$ of the second
integral in this equation. The non singularity of this limit
follows from the following argument. First, examine
$\lim_{\delta\rightarrow0}\Delta\phi(x)$ for a particle with a
different world line denoted by $z'(\tau')$, and defined such
that it coincides with the original world line $z(\tau)$ for
$\tau\le\tau_1$, and is slightly displaced with respect to
$z(\tau)$ for $\tau>\tau_1$, such that $z_0$ is not on $z'$. As
before, we assume that for the world line $z'(\tau')$ the
integrals in Eq. (\ref{phix},\ref{phixm}) converge in some local
neighborhood of the
 $z'(\tau')$, but not on $z'(\tau')$ itself.
 Since $z_0$ is not on
$z'(\tau')$, we find that for the world line $z'(\tau')$, $\Delta
\phi(z_0)$ is finite.
 This implies that by expressing $\Delta\phi(z_0)$ [or $\lim_{\delta\rightarrow0}\Delta\phi(x)$]
  by  Eq. (\ref{delphint});
  the second
integral in this equation is finite for the world line $z'(\tau')$,
and therefore it is also finite for the original world line
$z(\tau)$. We therefore conclude that $\Delta\phi(x)$ is
finite at the limit $\delta\rightarrow0$. Similarly, one can show that
${\Delta F}_{\mu}(x)$ is finite at the limit $\delta\rightarrow
0$; this limit, however, has directional irregularity (it depends on
$n_\mu$). This property is discussed in section \ref{massapp}.

%%%%%%%%%%%%%%%%%%%%%%%%%%%%%%%%%%%%%%% APPENDIX D %%%%%%%%%%%%%%%%%%%%%%%%%%%%%%%%%%%%%%%%%%%%%%%
\section{Calculation of the fourth term in de Sitter spacetime}\label{dsitter}

Here, we provide an example of the calculation of the fourth
term. We calculate the limit $m\rightarrow\infty$ of Eq.\
(\ref{fterm}) in de Sitter spacetime. In this case the retarded
Green's function of a massive scalar field is given by \cite{massg}
\begin{equation}\label{desitterg}
G_m(x|x')=\Theta(\Sigma(x),x')\left\{\left[\frac{\lambda
s}{\sinh(\lambda s)}\right]^{3/2}\delta(\sigma)
+\lambda^2P'_\eta[\cosh(\lambda s )]\Theta(-\sigma)\right\}.
\end{equation}
Here $\sigma=\sigma(x'|x)$, the Riemann tensor is given by
$R_{\mu\nu\eta\kappa}=\lambda^2(g_{\mu\eta}g_{\nu\kappa}-g_{\mu\kappa}g_{\nu\eta})$,
 $P'_\eta$ is the derivative of the Legendre function, and the order
$\eta$ satisfies the equation
$\eta(\eta+1)=2-\frac{m^2}{\lambda^2}$ (Note that $P_\eta$ is the
same for both roots of this equation.). We consider the case where
$\frac{m^2}{\lambda^2}>>2$, therefore $\eta\approx -\frac{1}{2}\pm
i\frac{m}{\lambda}$. Substituting the above expression for $G_m$
in Eq.\ (\ref{fterm}) gives
\begin{equation}\label{dsft}
\int_{-\infty}^{\tau_1+\epsilon}h_2{G_{m,}}_{\mu}(z_0|z(\tau))d\tau=
\lambda^2\int_{-\infty}^{\tau_1+\epsilon}h_2 \partial_\mu
\{P'_\eta[\cosh(\lambda s )]\}d\tau\,.
\end{equation}
Here, $s=s(z_0|z(\tau))$. Changing the integration variable to
$s$, and substituting $\partial_\mu
P'_\eta=s_{,\mu}\partial_s(P'_\eta)$ in Eq. (\ref{dsft}) and
integrating by parts gives
\begin{equation}\label{dsft2}
\int_{-\infty}^{\tau_1+\epsilon}h_2{G_{m,}}_{\mu}(z_0|z(\tau))d\tau=
-\lambda^2\int_{\infty}^{\tilde{s}_1}\partial_s\left(h_2s_{,\mu}\frac{d\tau}{ds}\right)
P'_\eta[\cosh(\lambda s )]ds\,.
\end{equation}
Here $\tilde{s}_1\equiv s(z_0|z(\tau_1+\epsilon))$. Note that both
boundary terms vanished since $h_2(\tilde{s}_1)=0$, and
 the asymptotic form of $P_\eta$ (as $s\rightarrow \infty$) \cite {GR} implies
 that asymptotically  $P'_\eta[\cosh(\lambda s)]\sim e^{(-3/2)\lambda s}$.
 We substitute \[
 P'_\eta[\cosh(\lambda s)]={[\lambda\sinh(\lambda s)]}^{-1}\frac{d}{ds}P_\eta\]
in Eq. (\ref{dsft2}), and integrate by parts once more. In the
resultant equation we substitute the following expression for
the Legendre function $P_\eta$ \cite{GR}:
\[
P_\eta[\cosh(\lambda s)]=\frac{1}{\sqrt{\pi}}\left[
\frac{\Gamma(\eta+1/2)}{\Gamma(\eta+1)}
 \frac{e^{(\eta+1)\lambda s}}{{(e^{2\lambda s}-1)}^{1/2}}
F\left(\frac{1}{2},\frac{1}{2},\frac{1}{2}-\eta,{(1-e^{2\lambda
 s})}^{-1}\right)+ c.c.\right]\,,
\]
where $F$ is the hypergeometric function. Noting that
asymptotically $\frac{\Gamma(z)}{\Gamma(z+1/2)}\sim z^{-1/2}$, we
obtain
\[
\lim_{m\rightarrow\infty}\int_{-\infty}^{\tau_1+\epsilon}h_2{G_{m,}}_{\mu}(z_0|z(\tau))d\tau\
=0\,,\] which conforms with our previous result.


\begin{thebibliography}{99}
\bibitem{Gdec}
The massless scalar Green's function is given in \cite{DB}, both
massless and massive scalar Green's functions (as well as more
general Green's functions) are given by theorem 4.3.1 in
\cite{Friedlander} .
\bibitem{massg}
The 'tail' part of this Green's function is given in
\cite{Friedlander}, equation (4.6.16). The 'direct' part of this
Green's function is not required here, and its explicit expression
is given only for completeness. The calculation of this dircet
part is obtained by integrating the differential equation for $U$
given in \cite{DB}, the detailed calculation is omitted.
\end{thebibliography}
\end{document}